\documentclass[reprint,aps,pra,superscriptaddress,notitlepage]{revtex4-2}
\usepackage{amssymb,amsmath}
\usepackage{graphicx}
\usepackage{dcolumn}
\usepackage{multirow}
\usepackage[english]{babel}
\usepackage{bm}
\usepackage[utf8]{inputenc}
\usepackage[T1]{fontenc}
\usepackage{mathptmx}
\usepackage{etoolbox}
\usepackage{braket}
\usepackage[colorlinks=true, linkcolor=black, citecolor=blue, urlcolor=blue]{hyperref}
\usepackage{xr}
\usepackage{color}
\usepackage{color,soul} 
\sethlcolor{white}
\soulregister\cite7
\soulregister\ref7
\usepackage{textcomp}

\begin{document}

\def\simlt{\mathrel{\lower .3ex \rlap{$\sim$}\raise .5ex \hbox{$<$}}}

\title{\textbf{\fontfamily{phv}\selectfont 
Single-shot latched readout of a quantum dot qubit using barrier gate pulsing}}
\author{Sanghyeok Park}
\thanks{These authors contributed equally to this work.}
\affiliation{Department of Physics, University of Wisconsin-Madison, Madison, WI 53706, USA}
\author{Jared Benson}
\thanks{These authors contributed equally to this work.}
\affiliation{Department of Physics, University of Wisconsin-Madison, Madison, WI 53706, USA}
\author{J. Corrigan}
\thanks{Present address: Intel Components Research, Intel Corporation, Hillsboro, OR 97124, USA.}
\affiliation{Department of Physics, University of Wisconsin-Madison, Madison, WI 53706, USA}
\author{J. P. Dodson}
\thanks{Present address: HRL Laboratories, LLC., 3011 Malibu Canyon Road, Malibu, CA 90265, USA.}
\affiliation{Department of Physics, University of Wisconsin-Madison, Madison, WI 53706, USA}
\author{S. N. Coppersmith}
\affiliation{Department of Physics, University of Wisconsin-Madison, Madison, WI 53706, USA}
\affiliation{School of Physics, University of New South Wales, Sydney, NSW 2052, Australia}
\author{Mark Friesen}
\affiliation{Department of Physics, University of Wisconsin-Madison, Madison, WI 53706, USA}
\author{M. A. Eriksson}
\affiliation{Department of Physics, University of Wisconsin-Madison, Madison, WI 53706, USA}
\email{maeriksson@wisc.edu}

\begin{abstract}
Latching techniques are widely used to enhance readout of qubits. These methods require precise tuning of multiple tunnel rates, which can be challenging to achieve under realistic experimental conditions, such as when a qubit is coupled to a single reservoir. Here, we present a method for single-shot measurement of a quantum dot qubit with a single reservoir using a latched-readout scheme. Our approach involves pulsing a barrier gate to dynamically control qubit-to-reservoir tunnel rates, a method that is readily applicable to the latched readout of various spin-based qubits. We use this method to enable qubit state latching and to reduce the qubit reset time in measurements of coherent Larmor oscillations of a Si/SiGe quantum dot hybrid qubit.
\end{abstract}
\maketitle

\section*{Introduction}
Single-shot readout, an essential technique for many quantum control protocols, has been reported for a variety of spin-based semiconductor quantum dot qubits~\cite{elzerman2004single, hanson2005single, barthel2009rapid,morello2010single,medford2013self,doi:10.1021/acs.nanolett.1c00783}. Latched readout improves on this technique by mapping short-lived qubit states to metastable charge states~\cite{studenikin2012enhanced, mason2015role, PhysRevLett.119.017701, PhysRevX.8.021046, PhysRevLett.127.127701}.
While latching mitigates problems arising from fast $T_1$ relaxation following spin-to-charge conversion during readout, it imposes stringent requirements on the tunnel rates~\cite{corrigan2023latched}. These conditions can be difficult to realize simultaneously, particularly when coupling to a single reservoir. This strategy is useful for scaling up qubit numbers, but often causes tunnel rates out of the latched state that are too fast for single-shot measurement.

Here, we report a baseband barrier-gate pulsing scheme that facilitates latching with a single reservoir. In such an approach, without a second reservoir the inter-dot tunnel rate is important in determining the effective tunnel rate to the dot that is not directly coupled to a reservoir. However, this inter-dot tunnel rate, also must be carefully set to enable control of the qubit itself. To solve this problem, we dynamically control the gate voltage that determines the tunnel rate to the single reservoir, which enables single-shot latched readout without the need for two reservoirs. This method can be readily adapted to any type of quantum dot qubit that employs latched readout. Here we demonstrate coherent Larmor oscillations of a quantum dot hybrid qubit (QDHQ)~\cite{shi_fast_2012} using this latching scheme to perform single-shot readout.

\section*{Results}
\begin{figure}[h!]
\includegraphics[width=\columnwidth]{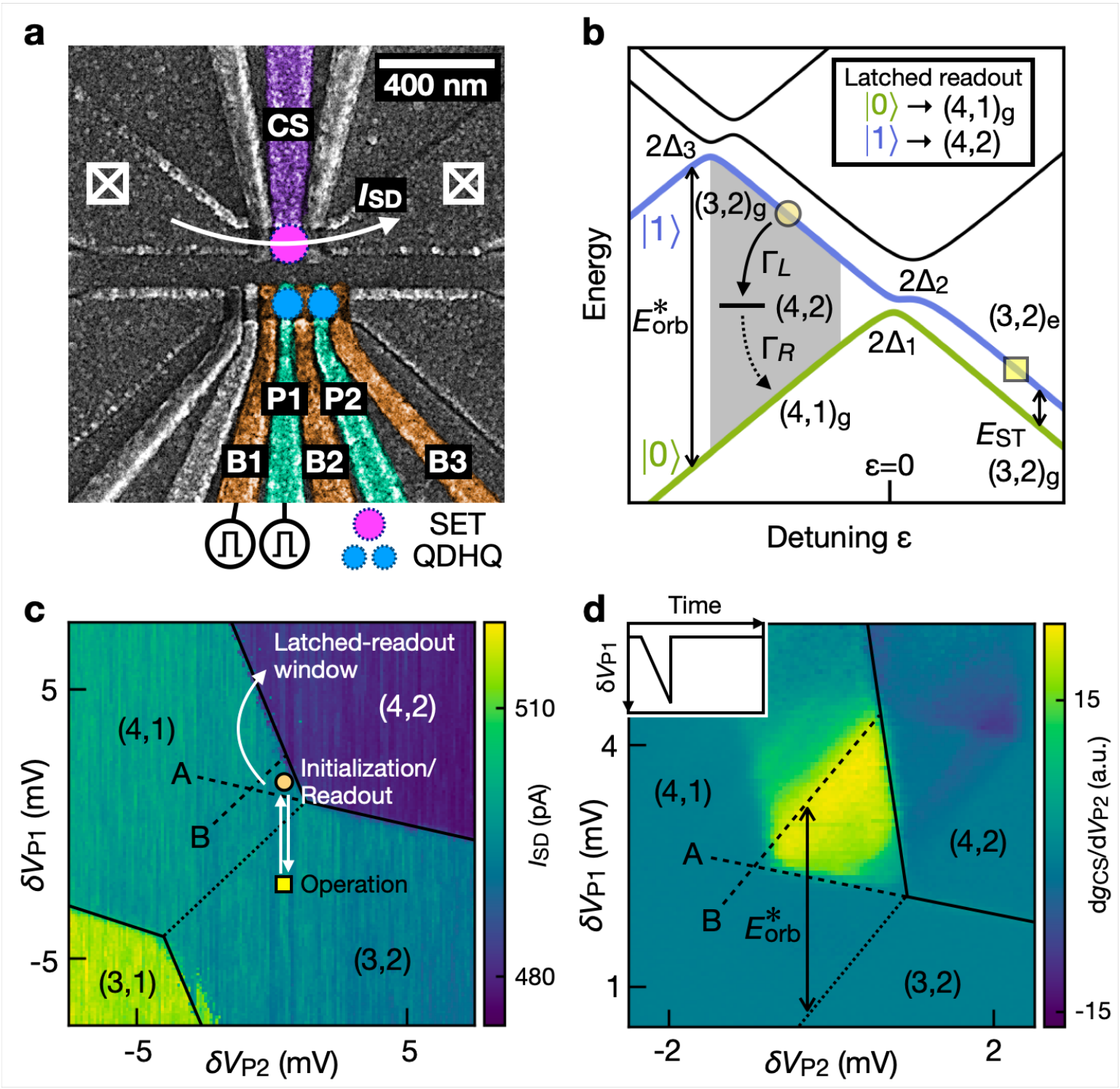}
\caption{\label{fig:fig1} Device layout and latched readout mechanism for the quantum dot hybrid qubit (QDHQ). \textbf{a} False-colored SEM micrograph of a device lithographically identical to the one under study. \textbf{b} Energy dispersion diagram for the 5-electron QDHQ. The latched readout mechanism via the (4,2) charge state is represented in the gray region. $\Gamma_L$ and $\Gamma_R$ are the tunnel rates of the P1 and P2 dots, respectively. $E^*_{\text{orb}}$ is the multi-electron orbital splitting of the P1 dot, and $E_{ST}$ is the single-triplet splitting of the P2 dot. $2\Delta_1$, $2\Delta_2$, and $2\Delta_3$ are the splittings at the anti-crossings between the (4,1) and (3,2) charge states. \textbf{c} Latched readout window of the QDHQ on the charge stability diagram, enclosed by dashed lines A and B. Line A extends from the (4,2)-(3,2) transition line, and line B is energetically positioned $E^*_{\text{orb}}$ apart from the polarization line. \textbf{d} Latching signal within the readout window, acquired with time-averaged measurement using a lock-in amplifier while applying a triangular latching pulse (inset) to gate P1.}
\end{figure}

Figure~\ref{fig:fig1} shows the device layout and the operational scheme for latched readout of a QDHQ. Figure~\ref{fig:fig1}a is a false-color scanning electron microscope (SEM) image of an overlapping-gate device fabricated on a Si/SiGe heterostructure similar to that measured here~\cite{Dodson:2020p505001}. Gate CS (purple) hosts a charge sensor (magenta circle) that is used for qubit measurement. A double dot (blue circles) is formed under plunger gates P1 and P2 (green), and the device tunnel couplings are tuned by barrier gates B1, B2, and B3 (orange). The left electron reservoir of the qubit is extended up to gate B1. High frequency lines are connected to gates P1 and B1, allowing for rapid control of the double-dot detuning $\varepsilon$ and the left dot tunnel rate, respectively. See Methods for more details of the experimental setup.

Figure~\ref{fig:fig1}b presents the energy dispersion diagram of the QDHQ in the $(N_{\text{P1}}, N_{\text{P2}}) \rightarrow$ (4,1)-(3,2) regime. The green and blue curves are the logical qubit states $\ket{0}$ and $\ket{1}$, respectively. In this scheme, the detuning point marked by the yellow square is used for qubit manipulation, where the qubit energy splitting is $E_{\text{ST}}$, the singlet-triplet splitting of the P2 dot. The gray-shaded region indicates the latched-readout window, whose size is directly proportional to $E^*_{\text{orb}}$, the multi-electron orbital splitting of the P1 dot.

Successful latched readout requires the fulfillment of two critical conditions. First, for qubit state $\ket{1}$ in (3,2)$_{\text{g}}$ to decay into the metastable (4,2) charge state, we must have
\begin{equation}
    \renewcommand{\theequation}{Condition~\arabic{equation}}
    \Gamma_L \gg T_1^{-1},
    \label{condition1}
\end{equation}
where $\Gamma_L$ is the left-dot (P1) tunnel rate, and $T_1^{-1}$ is the relaxation rate. Second, the right-dot (P2) tunnel rate, $\Gamma_R$, which determines the latch duration, must obey
\begin{equation}
    \renewcommand{\theequation}{Condition~\arabic{equation}}
    \Gamma_R \ll \Delta f_{BW},
    \label{condition2}
\end{equation}
where $\Delta f_{BW}$ is the measurement bandwidth. In this  work we show that these conditions can be met with a single reservoir on the left, achieved by completely pinching off gate B3. $\Gamma_R$ is then set by a cotunneling process through the left dot. With these conditions met, latching into (4,2) will occur because the relaxation process from $(3,2)_{\text{g}} \rightarrow (4,1)_{\text{g}}$ occurs in two steps involving this intermediate latching state (4,2) and facilitated by the fact that $\Gamma_L \gg \Gamma_R$.

Figure~\ref{fig:fig1}c shows the latched-readout window on a charge stability map, where the QDHQ is initialized and measured in the (4,1) charge configuration. Here, Line A is an extension of the (4,2)-(3,2) transition line, and line B is energetically separated by $E^*_{\text{orb}}$ from the polarization line. Since the (4,2) energy level must fall between the (4,1)$_{\text{g}}$ and (3,2)$_{\text{g}}$ energy levels to enable latching, the latched-readout window is bordered below by line A. Line B forms another boundary for the window, as it corresponds to the anti-crossing 2$\Delta_3$ depicted in Fig.~\ref{fig:fig1}b, where the (4,1) orbital excitation becomes involved in the process.

After initialization into the (4,1)$_{\text{g}}$ state, the QDHQ is pulsed into the (3,2) region for qubit operation. Figure~\ref{fig:fig1}d displays the latching signal acquired using time-averaged measurement of a lock-in amplifier with a triangular latching pulse (inset of Fig.~\ref{fig:fig1}d) applied to gate P1. This latching pulse adiabatically ramps the qubit across the polarization line and abruptly returns it to the readout window, thereby populating qubit state $\ket{1}$ and initiating the latching process. The measured latching signal matches the triangular readout window explained in Fig.~\ref{fig:fig1}c. In addition, some of the latching signal can also be measured beyond line B; however, this signal contains contributions from multiple processes other than simple latching and relies on $\Delta_3$.

\begin{figure*}[ht!]
\centering
\includegraphics[width=\textwidth]{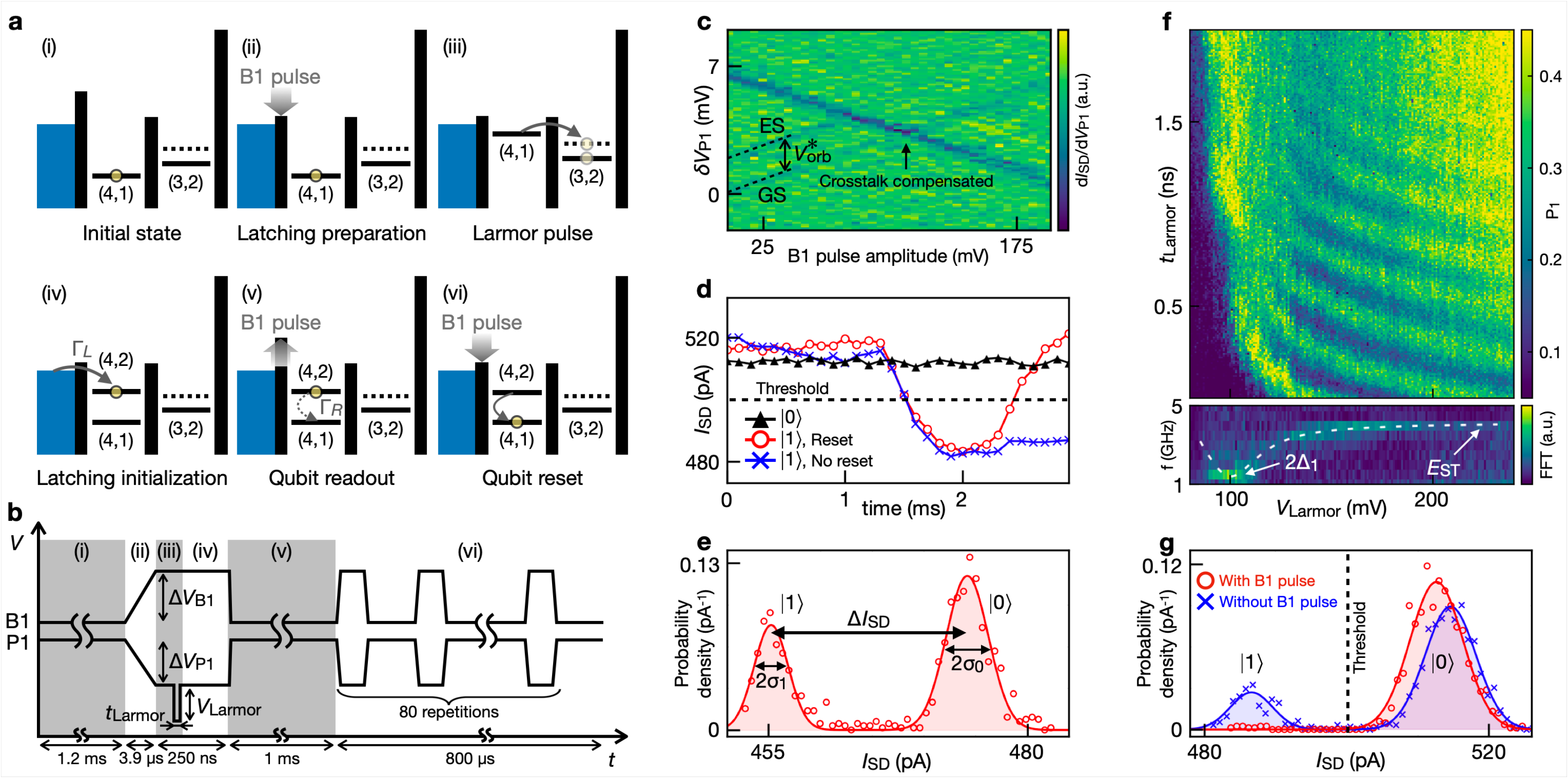}
\caption{\label{fig:fig2} A barrier gate pulsing scheme for single-shot latched readout and coherent Larmor oscillations of the QDHQ. \textbf{a} Diagram of the manipulation and readout process of the logical state $\ket{1}$ during the Larmor experiments, with the detunings are exaggerated for visual clarity. \textbf{b} Pulse shapes applied to gates P1 and B1 throughout the experiment stages (i) to (vi), as outlined in \textbf{a}. Note that the Larmor pulse in stages (iii) and (iv) is not to scale for demonstration purposes. \textbf{c} Pulsed-gate crosstalk measurement for gates P1 and B1. The arrow marks the point where the ground state (GS) electron-loading line intersects with the electron-unloading line, indicating that the effect of the P1 pulse on the QDHQ is fully canceled by the B1 pulse. \textbf{d} Example charge sensor time traces. The qubit state is identified as $\ket{1}$ when latching occurs and its signal trace surpasses a threshold. \textbf{e} Probability density plot for qubit latching, measured during qubit readout stage. Two peaks for logical states $\ket{0}$ and $\ket{1}$ are well separated with signal-to-noise ratio (SNR) of 10.2 and charge sensitivity of 3.10$\times 10^{-3} \text{e}/\sqrt{\text{Hz}}$. \textbf{f} Larmor oscillations of the QDHQ. Qubit parameters are extracted from its fast Fourier transform (FFT) displayed at the bottom: $\Delta_1 \approx$ 750 MHz and $E_{\text{ST}} \approx$ 4.0 GHz. The curve shown in the FFT data aligns with the energy splitting of the QDHQ (white dashed line). \textbf{g} Probability density plot for qubit reset, measured 2 ms after qubit readout. The initialization probabilities are ~98$\%$ with the reset pulse and ~80$\%$ without any reset pulse (1000 iterations each).}
\end{figure*}

In this conventional latched-readout scheme, \emph{single-shot measurement} remains challenging due to competing requirements during the manipulation and readout phases of the experiment. Crucially, as the qubit moves between the readout and operation points, it passes through anti-crossings, either adiabatically or at an intermediate rate to enable Landau-Zener (LZ) transitions. In both cases the inter-dot tunnel couplings $\Delta_1$ and $\Delta_2$ must be made reasonably large. If they are too small, either adiabatic pulses or LZ pulses will need to be so slow that decoherence will occur at detunings $\varepsilon$ that reside in between the $\Delta_1$ and $\Delta_2$ anticrossings shown in Fig.~1b. Having large $\Delta_1$ and $\Delta_2$ has an important effect: the right-dot tunnel rate $\Gamma_R$, which controls the lifetime of the (4,2) state used for latched readout, is dominated by a cotunneling process that increases as $\Delta_1$ increases~\cite{PhysRevB.78.155309, braakman2013long}. When $\Delta_1$ is increased to allow for fast, coherent manipulation, the latch reset rate increases, and it can become too fast to satisfy the latched readout Condition~\ref{condition2}. If the left-dot tunnel rate $\Gamma_L$ is decreased to mitigate this problem, it is then no longer possible to initialize the latch quickly enough, violating Condition~\ref{condition1}.

Here, we introduce a new method, using pulses applied to a single barrier gate to sequentially fulfill both latched readout conditions. As illustrated in Fig.~\ref{fig:fig2}a, with the timing shown in Fig.~\ref{fig:fig2}b, the process begins with the qubit initialized in the (4,1) ground state (i). A baseband pulse is then applied to gate B1 (ii), followed by a Larmor pulse on gate P1 that leaves the qubit in either the (3,2)$_{\text{g}}$ or (3,2)$_{\text{e}}$ state, depending on the time, $t_{\text{Larmor}}$, spent at the qubit operation point (iii) before returning to negative detuning. After this Larmor pulse, the qubit is transferred to either (4,1)$_{\text{g}}$ or (3,2)$_{\text{g}}$ in the readout window. If the qubit returns in the (3,2)$_{\text{g}}$ charge state, corresponding to qubit state $\ket{1}$, during step (iv) it latches into the (4,2) state, as illustrated in Fig.~\ref{fig:fig1}b. The B1 pulse is terminated shortly after the latch initializes to ensure the lifetime of the latched state is sufficiently long for readout (v). Finally, a sequence of pulses is applied to gate B1 to accelerate the rate the qubit resets to the initial (4,1)$_{\text{g}}$ state (vi). Note that when pulsing gate B1, there is significant crosstalk to the adjacent P1 dot that must be corrected to maintain the readout point of the qubit. Using the data shown in Fig.~\ref{fig:fig2}c, we determine the correct compensation pulse to apply to gate P1 to cancel this crosstalk. See Methods for a detailed description of the voltage pulses.

Using this method we show in Fig.~\ref{fig:fig2}d example single-shot traces corresponding to measurement of both $\ket{0}$ and $\ket{1}$. Figure~\ref{fig:fig2}e presents a probability density plot of the minimum $I_{\text{SD}}$ observed after both the triangular latching pulse and the barrier gate pulse are applied. It clearly shows two well-separated peaks, each representing a different logical qubit state. The probability density can be well described by a mixture of two Gaussian distributions, with $\Delta I_{SD}$, a difference in means, and variances of $\sigma_0^2$ and $\sigma_1^2$, respectively. The signal-to-noise ratio (SNR) for readout is given by $\Delta I_{\text{SD}}/\sqrt{\smash[b]{1/2(\sigma_0^2+\sigma_1^2)}} = 10.2$, and the charge sensitivity is determined to be $\text{e}/(\text{SNR} \cdot \sqrt{\smash[b]{\Delta f_{BW}}}) = 3.10\times 10^{-3} \text{e}/\sqrt{\text{Hz}}$.

The blue data in Fig.~\ref{fig:fig2}d demonstrate the importance of the reset pulses applied to gate B1. Without the reset pulses---stage~(vi) in Fig.~\ref{fig:fig2}a---the qubit persists for a long time in the latched state, whereas with the reset pulses, the latch resets rapidly. Fig.~\ref{fig:fig2}g shows a probability density plot for the initialization probabilities of the logical states, measured 2 ms after readout. Without applying any reset pulse (blue), the initialization probability stands at about 80\%. However, when a reset pulse is applied (red data), this probability already increases to 98\% at a time of 2~ms, representing a 15$\times$ speedup.

As a demonstration of the utility of the barrier-gate pulsing scheme, we report coherent Larmor oscillations of the QDHQ using single-shot latched readout in Fig.~\ref{fig:fig2}f, as a function of Larmor pulse amplitudes $V_{\text{Larmor}}$ and wait times $t_{\text{Larmor}}$. From the fast Fourier transform (FFT) of the Larmor data, also included in Fig.~\ref{fig:fig2}f, we observe both charge-like ($V_{\text{Larmor}} \approx$ 100 mV) and spin-like ($V_{\text{Larmor}} \approx$ 200 mV) features of the QDHQ, which have been discussed in previous work on hybrid qubits~\cite{shi_fast_2014, PhysRevLett.116.086801, kim_high-fidelity_2015}. To a good approximation, the minimum value of the qubit frequency (white dashed line) corresponds to $2\Delta_1/h = 1.5$ GHz. We also obtain the value of the singlet-triplet splitting of the P2 dot in the asymptotic regime, yielding $E_{\text{ST}}/h= 4.0$ GHz, consistent with the value measured via magneto-spectroscopy. (See Supplemental material.)

\section*{Discussion}

Latched readout can be used to read out the state of a qubit, when the measurement time exceeds the $T_1$ relaxation rate, through careful tuning of the quantum dot tunnel rates. We present a barrier gate pulsing method that improves latched readout by meeting its requirements sequentially, rather than simultaneously. We have shown that this technique enables coherent manipulation of a QDHQ measured with single-shot latched readout, and speeds up its reset process. To illustrate the importance of the latter point, consider a qubit experiment in which it takes 1 ms to read out the qubit state using latched readout. Based on simple rate equations, we estimate that, for the qubit to remain in the logical $\vert 1\rangle$ state for at least 99\% of the 1 ms measurement time, the time to tunnel out of the qubit state should be at least
\begin{equation}
    t_{\mathrm{out}} = -\frac{1\mathrm{ms}}{\ln{(0.99)}}\, = 99.5\,\mathrm{ms}.
    \label{t_out}
\end{equation}
Given this value of $t_\mathrm{out}$, the time required to initialize the qubit back into the logical $\vert 0\rangle$ state after measurement at least 99\% of the time is at least
\begin{equation}
    t_{\mathrm{reset}} = -t_\mathrm{out}\ln{(0.01)} = 458\,\mathrm{ms}.
    \label{t_out}
\end{equation}
This is an excessively long initialization time for a qubit, which would not be feasible for qubit manipulations, like the Larmor oscillation experiment shown in Fig.~\ref{fig:fig2}f. Note that the measurement bandwidth $\Delta f_{BW}$ in this experiment is 1 kHz, which can be further improved by using RF readout methods, such as RF-reflectometry or RF-SET with cryogenic amplification, making it possible to read out the qubit state faster. However, even with these faster measurement methods, qubit initialization will still benefit from the methods and type of improvement demonstrated here. For example, in an experiment in which the qubit can be read out in 1 \textmu s, the analogous reset time would be at least 458 \textmu s in the absence of a reset pulse sequence.

The technique described here will be relevant for many future experiments investigating coherent qubit manipulations. It is not limited to QDHQ but broadly applicable to any qubit whose readout scheme is based on spin-to-charge conversion in a double dot, including singlet-triplet qubits~\cite{petta2005coherent}, exchange-only qubits~\cite{laird_coherent_2010} and single-spin qubits in parity-readout mode~\cite{seedhouse_pauli_2021}.


\section*{Methods}
\subsection*{Experimental Setup}
The device under test is an overlapping-gate quantum dot device fabricated as in Ref.~\cite{Dodson:2020p505001}. The experiments are conducted in a dilution refrigerator with a base temperature $\lesssim 30$ mK. DC signals are supplied by SIM928 isolated voltage sources and RF signals are synthesized by a one-channel Tektronix AWG70001A arbitrary waveform generator (AWG) and a two-channel Tektronix AWG70002A AWG. DC and RF signals are combined using bias tees (Picosecond Pulse Labs 5546) mounted to the mixing chamber of the dilution refrigerator.

Charge sensing is performed with a sensor dot under gate CS whose current $I_{\text{SD}}$ is amplified at room temperature using a DL Instruments Model 1211 current preamplifier and an SRS SR560 voltage preamplifier. DC signals are digitized with an NI-USB 6216 (for $I_{\text{SD}}$ trace measurement) or a Quantum Machines OPX+ (for single-shot readout), and lock-in signals are measured with a Signal Recovery 7265 DSP lock-in amplifier.

\subsection*{Detailed Description of the Voltage Pulses}
Two distinct types of manipulation pulses are applied to the P1 plunger via RF lines in this experiment; a triangular latching pulse (illustrated in Fig.~\ref{fig:fig1}d) and a Larmor pulse (illustrated in Fig.~\ref{fig:fig2}b). The latching pulse is designed to test the latching process as described in the main text. It has a ramp-in duration of 2 ns and a ramp-out duration of 20 ps. The Larmor pulse is utilized to prepare the qubit as a superposition state of $\ket{0}$ and $\ket{1}$. Ramp times for the Larmor pulse are optimized to 400 ps, effectively acting as a $\pi$/2 rotation on the qubit Bloch sphere.

The crosstalk measurement in Fig.~\ref{fig:fig2}c resembles a typical pulsed-gate spectroscopy measurement~\cite{10.1063/1.1757023}. First, we apply a square pulse with an amplitude of 50 mV and a frequency of 100 kHz to gate P1. The action of this pulse is to split the usual (4,2)-(4,1) transition line into a pair of electron loading and unloading lines. Simultaneously, an in-phase square pulse with the same frequency is applied to gate B1. The crosstalk from gate B1 onto gate P1 effectively cancels some of the P1 pulse. By varying the amplitude of the B1 pulse, we find the crossover of the electron loading and unloading lines, which represents the pulse amplitude at which the crosstalk is effectively canceled. By comparing the amplitudes of the P1 and B1 pulses, we extract a conversion factor that we use to calibrate the compensation pulses used when pulsing the barrier gate voltage: $\Delta V_{\text{B1}}/\Delta V_{\text{P1}}$ = 2.24. Additionally, multiple electron loading lines appear when a square pulse with a large amplitude is applied to gate P1. We find the separation between these lines, calculated as $\alpha_{\text{P1, L}} V_{\text{orb}}^*$, corresponds to the multi-electron orbital splitting of the P1 dot $E_{\text{orb}}^*$, shown in Fig.~\ref{fig:fig1}d. Here, $\alpha_{\text{P1, L}}$ is the lever arm between P1 and the left dot.

For the barrier gate pulsing scheme, three different pulses are applied to gates P1 and B1: P1 Larmor pulse, B1 pulse, P1 compensation pulse. A one-channel AWG is used for the P1 Larmor pulse, while a two-channel AWG is used for the B1 pulse and the P1 compensation pulse. We use an `interleaved' function, which provides fine control of the offset between the signals from two internal channels, to sweep wait time t$_{\text{Larmor}}$ with picosecond time resolution~\cite{macquarrie_progress_2020}. Larmor pulses and compensation pulses are merged via a power splitter at room temperature. The phase offset from the power splitter is measured with an oscilloscope, and a time delay of 2 ns is applied to cancel out their offset. To avoid a potential timing mismatch between gates B1 and P1, we utilize the same length of phase-stable SMA cables for both connections. Note that the initial B1 pulse ramp occurs before the qubit manipulation pulse so that it can be made long enough to avoid the same problem on the scale of the unknown length mismatch between the P1 and B1 high frequency lines inside the dilution refrigerator. In Fig.~\ref{fig:fig2}b, $\Delta V_{\text{B1}}=250$ mV. The ramp-in duration in stage (ii) is set as 3.9 \textmu s to prevent issues from the aforementioned cable length mismatch, while ramp-out duration in stage (iv) is set to 5 ns to reduce any possible decoherence. 5 ns of wait time in stage (iv) and repetition number 80 of the B1 reset pulses in stage (vi) are carefully optimized and set for high latched readout probability. A single B1 reset pulse consists of two 500 ns ramps and a 3 \textmu s wait time. This is repeated in stage (vi) instead of a single pulse with a longer duration to enable the pulse to pass through the bias tee, whose cutoff frequency is 3.5 kHz.

\subsection*{Measurement}
For the time-averaged measurement in Fig.~\ref{fig:fig1}d, the lock-in amplifier is configured to compare the charge sensor current in the presence (absence) of the triangular latching pulse by synchronizing the AWG output on (off) state with the high (low) side of an external 217 Hz square pulse reference \cite{PhysRevLett.127.127701}.

Real-time thresholding for single-shot readout is performed with a Quantum Machines OPX+ which triggers the RF manipulation pulses from the AWGs using one of its digital output channels. During the single-shot readout in the barrier gate pulsing scheme, DC signals are measured for 600 \textmu s split across 6 measurements of 100 \textmu s with a sample rate of 1 GS/s at stage (v) in Fig.~\ref{fig:fig2}b. If at least one of the measurements of $I_{\text{SD}}$ passes the threshold, the qubit state is labeled as the logical state $\ket{1}$. The OPX+ has a DC offset at its analog input port. This results in an offset of around 245 mV to the NI-USB 6216 input during the readout, which is removed from the $I_{\text{SD}}$ values in Fig.~\ref{fig:fig2}d by subtracting this constant offset from each data point.

\section*{Data Availability}
The datasets generated during the current study are available in a \href{https://doi.org/10.5281/zenodo.13368027}{Zenodo repository}~\cite{zenodo}.

\section*{Acknowledgements}
Research was sponsored in part by the Department of Defense and the Army Research Office (ARO) under Grant Numbers W911NF-17-1-0274 and W911NF-23-1-0110. We acknowledge the use of facilities supported by NSF through the UW-Madison MRSEC (DMR-2309000) and the NSF MRI program (DMR-1625348). The authors thank HRL for support and L. F. Edge of HRL Laboratories for providing the Si/SiGe heterostructure used in this work. The views and conclusions contained in this document are those of the authors and should not be interpreted as representing the official policies, either expressed or implied, of the Army Research Office (ARO), or the U.S. Government. The U.S. Government is authorized to reproduce and distribute reprints for Government purposes notwithstanding any copyright notation herein.

\section*{Author Contributions}
SP, JB performed the measurements. JC and JPD contributed to the experimental setup and methods. JPD fabricated the sample. SP, JB, SNC, MF, MAE analyzed the data. All the authors wrote the manuscript.

\section*{Competing Interests}
The authors declare no competing interests.
\vspace{1em}

\section*{Supplemental material for `Single-shot latched readout of a quantum dot qubit using barrier gate pulsing'}

\subsection*{Determining Lever-arms} 
The gate-to-dot lever arm is obtained by performing magneto-spectroscopy for the P1 dot. Other lever arms are calculated from the slopes of the transition lines and polarization line in the charge stability diagram based on Ref.~\cite{house2011non} and the approach presented in the supplemental material of Ref.~\cite{corrigan2023longitudinal}.

\setlength{\tabcolsep}{0.5em}
{\renewcommand{\arraystretch}{1.2}
\begin{table}[h!]
    \caption{Gate lever arms}
    \begin{tabular}{c|c}
        \toprule
        Lever arms & Values~(eV/V)\\
        \colrule
        $\alpha_{\mathrm{P1}, L}$ & 0.1983 $\pm$ 0.0002\\
        $\alpha_{\mathrm{P2}, L}$ & 0.0378 $\pm$ 0.0048\\
        $\alpha_{\mathrm{P1}, R}$ & 0.0458 $\pm$ 0.0049\\
        $\alpha_{\mathrm{P2}, R}$ & 0.2750 $\pm$ 0.0457\\
        $\alpha_{\mathrm{P1}, \varepsilon}$ & 0.1525 $\pm$ 0.0049\\
        \botrule
    \end{tabular}
\end{table}
}

\subsection*{Measuring Qubit Parameters}
Pulsed-gate spectroscopy on the (3,2)-(4,2) transition line is used to determine $E^*_{\text{orb}}$ of the P1 dot. Magneto-spectroscopy on the (4,1)-(4,2) transition line is used to determine $E_{\text{ST}}$ of the P2 dot. $T_e$ is calculated based on the broadening of the transition line. $T_1$ is measured by using the latched readout outside the triangle region in Fig.~\ref{fig:fig1}c and observing the signal decay as a function of wait time the state stays inside the triangle region. Tunnel rates are measured by observing the tunnel events on the transition line and binning the time the electron spends on the dot.

\setlength{\tabcolsep}{0.5em}
{\renewcommand{\arraystretch}{1.2}
\begin{table}[h!]
    \caption{Device parameters}
    \begin{tabular}{c c c c}
        \toprule
        $E^*_{\text{orb}}$~(\textmu eV) & $E_{\text{ST}}$~(\textmu eV) & $T_1$~(ns) & $T_e$~(mK)\\
        \colrule
        416 $\pm$ 5 & 17 & 180-250 & 123\\
        \botrule
    \end{tabular}
\end{table}
}


\begin{thebibliography}{10}
\expandafter\ifx\csname url\endcsname\relax
  \def\url#1{\texttt{#1}}\fi
\expandafter\ifx\csname urlprefix\endcsname\relax\def\urlprefix{URL }\fi
\providecommand{\bibinfo}[2]{#2}
\providecommand{\eprint}[2][]{\url{#2}}

\bibitem{elzerman2004single}
\bibinfo{author}{Elzerman, J.~M.} \emph{et~al.}
\newblock \bibinfo{title}{{Single-shot read-out of an individual electron spin
  in a quantum dot}}.
\newblock \emph{\bibinfo{journal}{Nature}} \textbf{\bibinfo{volume}{430}},
  \bibinfo{pages}{431--435} (\bibinfo{year}{2004}).

\bibitem{hanson2005single}
\bibinfo{author}{Hanson, R.} \emph{et~al.}
\newblock \bibinfo{title}{{Single-Shot Readout of Electron Spin States in a
  Quantum Dot Using Spin-Dependent Tunnel Rates}}.
\newblock \emph{\bibinfo{journal}{Phys. Rev. Lett.}}
  \textbf{\bibinfo{volume}{94}}, \bibinfo{pages}{196802}
  (\bibinfo{year}{2005}).

\bibitem{barthel2009rapid}
\bibinfo{author}{Barthel, C.}, \bibinfo{author}{Reilly, D.~J.},
  \bibinfo{author}{Marcus, C.~M.}, \bibinfo{author}{Hanson, M.~P.} \&
  \bibinfo{author}{Gossard, A.~C.}
\newblock \bibinfo{title}{{Rapid Single-Shot Measurement of a Singlet-Triplet
  Qubit}}.
\newblock \emph{\bibinfo{journal}{Phys. Rev. Lett.}}
  \textbf{\bibinfo{volume}{103}}, \bibinfo{pages}{160503}
  (\bibinfo{year}{2009}).

\bibitem{morello2010single}
\bibinfo{author}{Morello, A.} \emph{et~al.}
\newblock \bibinfo{title}{{Single-shot readout of an electron spin in
  silicon}}.
\newblock \emph{\bibinfo{journal}{Nature}} \textbf{\bibinfo{volume}{467}},
  \bibinfo{pages}{687--691} (\bibinfo{year}{2010}).

\bibitem{medford2013self}
\bibinfo{author}{Medford, J.} \emph{et~al.}
\newblock \bibinfo{title}{{Self-consistent measurement and state tomography of
  an exchange-only spin qubit}}.
\newblock \emph{\bibinfo{journal}{Nature Nanotechnology}}
  \textbf{\bibinfo{volume}{8}}, \bibinfo{pages}{654--659}
  (\bibinfo{year}{2013}).

\bibitem{doi:10.1021/acs.nanolett.1c00783}
\bibinfo{author}{Jang, W.} \emph{et~al.}
\newblock \bibinfo{title}{{Single-Shot Readout of a Driven Hybrid Qubit in a
  GaAs Double Quantum Dot}}.
\newblock \emph{\bibinfo{journal}{Nano Letters}} \textbf{\bibinfo{volume}{21}},
  \bibinfo{pages}{4999--5005} (\bibinfo{year}{2021}).

\bibitem{studenikin2012enhanced}
\bibinfo{author}{Studenikin, S.~A.} \emph{et~al.}
\newblock \bibinfo{title}{{Enhanced charge detection of spin qubit readout via
  an intermediate state}}.
\newblock \emph{\bibinfo{journal}{Applied Physics Letters}}
  \textbf{\bibinfo{volume}{101}}, \bibinfo{pages}{233101}
  (\bibinfo{year}{2012}).

\bibitem{mason2015role}
\bibinfo{author}{Mason, J.~D.} \emph{et~al.}
\newblock \bibinfo{title}{{Role of metastable charge states in a quantum-dot
  spin-qubit readout}}.
\newblock \emph{\bibinfo{journal}{Phys. Rev. B}} \textbf{\bibinfo{volume}{92}},
  \bibinfo{pages}{125434} (\bibinfo{year}{2015}).

\bibitem{PhysRevLett.119.017701}
\bibinfo{author}{Nakajima, T.} \emph{et~al.}
\newblock \bibinfo{title}{{Robust Single-Shot Spin Measurement with 99.5\%
  Fidelity in a Quantum Dot Array}}.
\newblock \emph{\bibinfo{journal}{Phys. Rev. Lett.}}
  \textbf{\bibinfo{volume}{119}}, \bibinfo{pages}{017701}
  (\bibinfo{year}{2017}).

\bibitem{PhysRevX.8.021046}
\bibinfo{author}{Harvey-Collard, P.} \emph{et~al.}
\newblock \bibinfo{title}{{High-Fidelity Single-Shot Readout for a Spin Qubit
  via an Enhanced Latching Mechanism}}.
\newblock \emph{\bibinfo{journal}{Phys. Rev. X}} \textbf{\bibinfo{volume}{8}},
  \bibinfo{pages}{021046} (\bibinfo{year}{2018}).

\bibitem{PhysRevLett.127.127701}
\bibinfo{author}{Corrigan, J.} \emph{et~al.}
\newblock \bibinfo{title}{{Coherent Control and Spectroscopy of a Semiconductor
  Quantum Dot Wigner Molecule}}.
\newblock \emph{\bibinfo{journal}{Phys. Rev. Lett.}}
  \textbf{\bibinfo{volume}{127}}, \bibinfo{pages}{127701}
  (\bibinfo{year}{2021}).

\bibitem{corrigan2023latched}
\bibinfo{author}{Corrigan, J.} \emph{et~al.}
\newblock \bibinfo{title}{{Latched readout for the quantum dot hybrid qubit}}.
\newblock \emph{\bibinfo{journal}{Applied Physics Letters}}
  \textbf{\bibinfo{volume}{122}}, \bibinfo{pages}{074001}
  (\bibinfo{year}{2023}).

\bibitem{shi_fast_2012}
\bibinfo{author}{Shi, Z.} \emph{et~al.}
\newblock \bibinfo{title}{{Fast Hybrid Silicon Double-Quantum-Dot Qubit}}.
\newblock \emph{\bibinfo{journal}{Phys. Rev. Lett.}}
  \textbf{\bibinfo{volume}{108}}, \bibinfo{pages}{140503}
  (\bibinfo{year}{2012}).

\bibitem{Dodson:2020p505001}
\bibinfo{author}{Dodson, J.~P.} \emph{et~al.}
\newblock \bibinfo{title}{{Fabrication process and failure analysis for robust
  quantum dots in silicon}}.
\newblock \emph{\bibinfo{journal}{Nanotechnology}}
  \textbf{\bibinfo{volume}{31}}, \bibinfo{pages}{505001}
  (\bibinfo{year}{2020}).

\bibitem{PhysRevB.78.155309}
\bibinfo{author}{Gustavsson, S.} \emph{et~al.}
\newblock \bibinfo{title}{{Detecting single-electron tunneling involving
  virtual processes in real time}}.
\newblock \emph{\bibinfo{journal}{Phys. Rev. B}} \textbf{\bibinfo{volume}{78}},
  \bibinfo{pages}{155309} (\bibinfo{year}{2008}).

\bibitem{braakman2013long}
\bibinfo{author}{Braakman, F.~R.}, \bibinfo{author}{Barthelemy, P.},
  \bibinfo{author}{Reichl, C.}, \bibinfo{author}{Wegscheider, W.} \&
  \bibinfo{author}{Vandersypen, L. M.~K.}
\newblock \bibinfo{title}{{Long-distance coherent coupling in a quantum dot
  array}}.
\newblock \emph{\bibinfo{journal}{Nature Nanotechnology}}
  \textbf{\bibinfo{volume}{8}}, \bibinfo{pages}{432--437}
  (\bibinfo{year}{2013}).

\bibitem{shi_fast_2014}
\bibinfo{author}{Shi, Z.} \emph{et~al.}
\newblock \bibinfo{title}{{Fast coherent manipulation of three-electron states
  in a double quantum dot}}.
\newblock \emph{\bibinfo{journal}{Nature Communications}}
  \textbf{\bibinfo{volume}{5}}, \bibinfo{pages}{3020} (\bibinfo{year}{2014}).

\bibitem{PhysRevLett.116.086801}
\bibinfo{author}{Cao, G.} \emph{et~al.}
\newblock \bibinfo{title}{{Tunable Hybrid Qubit in a GaAs Double Quantum Dot}}.
\newblock \emph{\bibinfo{journal}{Phys. Rev. Lett.}}
  \textbf{\bibinfo{volume}{116}}, \bibinfo{pages}{086801}
  (\bibinfo{year}{2016}).

\bibitem{kim_high-fidelity_2015}
\bibinfo{author}{Kim, D.} \emph{et~al.}
\newblock \bibinfo{title}{{High-fidelity resonant gating of a silicon-based
  quantum dot hybrid qubit}}.
\newblock \emph{\bibinfo{journal}{npj Quantum Information}}
  \textbf{\bibinfo{volume}{1}}, \bibinfo{pages}{15004} (\bibinfo{year}{2015}).

\bibitem{petta2005coherent}
\bibinfo{author}{Petta, J.~R.} \emph{et~al.}
\newblock \bibinfo{title}{{Coherent Manipulation of Coupled Electron Spins in
  Semiconductor Quantum Dots}}.
\newblock \emph{\bibinfo{journal}{Science}} \textbf{\bibinfo{volume}{309}},
  \bibinfo{pages}{2180--2184} (\bibinfo{year}{2005}).

\bibitem{laird_coherent_2010}
\bibinfo{author}{Laird, E.~A.} \emph{et~al.}
\newblock \bibinfo{title}{{Coherent spin manipulation in an exchange-only
  qubit}}.
\newblock \emph{\bibinfo{journal}{Phys. Rev. B}} \textbf{\bibinfo{volume}{82}},
  \bibinfo{pages}{075403} (\bibinfo{year}{2010}).

\bibitem{seedhouse_pauli_2021}
\bibinfo{author}{Seedhouse, A.~E.} \emph{et~al.}
\newblock \bibinfo{title}{{Pauli Blockade in Silicon Quantum Dots with
  Spin-Orbit Control}}.
\newblock \emph{\bibinfo{journal}{PRX Quantum}} \textbf{\bibinfo{volume}{2}},
  \bibinfo{pages}{010303} (\bibinfo{year}{2021}).

\bibitem{10.1063/1.1757023}
\bibinfo{author}{Elzerman, J.~M.}, \bibinfo{author}{Hanson, R.},
  \bibinfo{author}{Willems~van Beveren, L.~H.}, \bibinfo{author}{Vandersypen,
  L. M.~K.} \& \bibinfo{author}{Kouwenhoven, L.~P.}
\newblock \bibinfo{title}{{Excited-state spectroscopy on a nearly closed
  quantum dot via charge detection}}.
\newblock \emph{\bibinfo{journal}{Applied Physics Letters}}
  \textbf{\bibinfo{volume}{84}}, \bibinfo{pages}{4617--4619}
  (\bibinfo{year}{2004}).

\bibitem{macquarrie_progress_2020}
\bibinfo{author}{MacQuarrie, E.~R.} \emph{et~al.}
\newblock \bibinfo{title}{{Progress toward a capacitively mediated CNOT between
  two charge qubits in Si/SiGe}}.
\newblock \emph{\bibinfo{journal}{npj Quantum Information}}
  \textbf{\bibinfo{volume}{6}}, \bibinfo{pages}{81} (\bibinfo{year}{2020}).

\bibitem{zenodo}
\bibinfo{author}{Park, S.} \emph{et~al.}
\newblock \bibinfo{title}{{Source data for the publication "Single-shot latched
  readout of a quantum dot qubit using barrier gate pulsing"}}.
\newblock \emph{\bibinfo{journal}{Zenodo}}  (\bibinfo{year}{2024}).
\newblock
  \bibinfo{note}{\href{https://doi.org/10.5281/zenodo.13368027}{10.5281/zenodo.13368026}}.

\bibitem{house2011non}
\bibinfo{author}{House, M.~G.}, \bibinfo{author}{Pan, H.},
  \bibinfo{author}{Xiao, M.} \& \bibinfo{author}{Jiang, H.~W.}
\newblock \bibinfo{title}{{Non-equilibrium charge stability diagrams of a
  silicon double quantum dot}}.
\newblock \emph{\bibinfo{journal}{Applied Physics Letters}}
  \textbf{\bibinfo{volume}{99}}, \bibinfo{pages}{112116}
  (\bibinfo{year}{2011}).

\bibitem{corrigan2023longitudinal}
\bibinfo{author}{Corrigan, J.} \emph{et~al.}
\newblock \bibinfo{title}{{Longitudinal coupling between a
  ${\mathrm{Si}/\mathrm{Si}}_{1\ensuremath{-}x}{\mathrm{Ge}}_{x}$ double
  quantum dot and an off-chip $\mathrm{Ti}\mathrm{N}$ resonator}}.
\newblock \emph{\bibinfo{journal}{Phys. Rev. Appl.}}
  \textbf{\bibinfo{volume}{20}}, \bibinfo{pages}{064005}
  (\bibinfo{year}{2023}).

\end{thebibliography}
\end{document}